# A New Route to Fluorescent SWNT/Silica Nanocomposites: Balancing Fluorescence Intensity and Environmental Sensitivity

Juan G. Duque[1,†], Gautam Gupta[2,†], Laurent Cognet[3], Brahim Lounis[3], Stephen K. Doorn[2], and Andrew M. Dattelbaum[2*]

[1]Chemistry Division, Physical Chemistry and Applied Spectroscopy Group, MS J536, Los Alamos National Laboratory, Los Alamos, NM 87545

[2]Center for Integrated Nanotechnologies, MS K771, Los Alamos National Laboratory, Los Alamos, NM 87545

[3] Laboratoire Photonique Numérique et Nanosciences, Université de Bordeaux, Institut d'Optique Graduate School and CNRS, 33405 Talence, France

[*]To whom correspondence should be addressed: amdattel@lanl.gov
[†]Authors contributed equally to this work.

**ABSTRACT**

We investigate the relationship between photoluminescence (PL) intensity and environmental sensitivity of surfactant-wrapped single walled carbon nanotubes (SWNTs). SWNTs were studied under a variety of conditions in suspension as well as encapsulated in silica nanocomposites, which were prepared by an efficient chemical vapor into liquids (CViL) sol-gel process. The dramatically improved silica encapsulation process described here has several advantages, including fast preparation and high SWNT loading concentration, over other encapsulation methods used to prepare fluorescent SWNT/silica nanocomposites. Further, addition of glycerol to SWNT suspensions prior to performing the CViL sol-gel process allows for the preparation of freestanding fluorescent silica xerogels, which to the best of our knowledge is the first report of such nanocomposites. Our spectroscopic data on SWNTs suspended in aqueous surfactants or encapsulated in silica show that achieving maximum PL intensity results in decreased sensitivity of SWNT emission response to changes imparted by the local environment. In addition, silica encapsulation can be used to "lock-in" a surfactant micelle structure surrounding SWNTs to minimize interactions between SWNTs and ions/small molecules. Ultimately, our work demonstrates that one should consider a balance between maximum PL intensity and the ability to sense environmental changes when designing new SWNT systems for future sensing applications.

# INTRODUCTION

Single walled carbon nanotubes (SWNTs) possess unique mechanical, electrical and luminescence properties due to their nanoscale structure based on a cylindrical sheet of graphene.[1,2] These properties can be exploited by individually-suspending the SWNTs in aqueous solutions predominately via addition of surfactants.[3,4] In particular, individualized semiconducting SWNTs have been shown to exhibit near-infrared (NIR) photoluminescence (PL) that is dependent on SWNT chirality.[3,4] The PL signal from SWNTs has been shown to be important in a variety of *in vitro* and *in vivo* sensing and imaging applications,[5-7] as well as for the preparation of new optoelectronic materials.[8] Further, the PL signal from semiconducting SWNTs has been used to study redox interactions of SWNTs with solubilized dye[9] and transition metal complexes[10] to elucidate a number of chirality selective reaction trends that may have important implications for sensors, electronic applications, and separations of SWNTs. For example, reversible fluorescence quenching studies have been reported for sensing biomolecular species with high selectivity and sensitivity.[11,12]

A number of surfactant systems have been used in attempts to optimize the photoluminescence intensity signal from dispersed SWNTs.[3,13-16] Ideally, for imaging and sensing applications a higher PL signal would increase the signal-to-noise ratio resulting in increased sensitivity, ultimately down to the single molecule sensitivity.[17] Recent work[15] has shown that the surfactant assembly surrounding the SWNTs can be tuned to increase the PL signal from SWNTs; however, this may also limit the interaction between the local environment and the SWNTs, which is potentially problematic for development of new sensors.

To fully exploit the unique NIR PL properties of SWNTs for future device applications, there is also a need to develop SWNT composite materials that may readily encapsulate isolated and fluorescent SWNTs. But, the fragility of the SWNT/surfactant assembly has limited the ability to incorporate fluorescent SWNTs into composite materials. Polymers such as gelatin,[18] agarose[17] or poly(vinyl alcohol)[19] have been successfully used to make fluorescent SWNT/polymer composites. However, these systems extrinsically change the local environment of the nanotubes, which in turn induce significant changes in spectral intensity and peak position. Moreover, because of the nature of the polymer matrix preparation and chemical structure, the composite may suffer from relatively low SWNT loading concentrations and/or poor thermal stability. In contrast, silica is an ideal material for encapsulation of fluorescent materials, as it is generally inert, thermally stable, optically transparent, and permeable. Silica can also be prepared from molecular precursors in solution through sol-gel processes that can take advantage of highly concentrated fluorescent SWNT aqueous suspensions. But, traditional sol-gel procedures[20] are performed in alcohol solvents at very acidic or basic pH conditions, which are conditions that can disrupt the SWNT/surfactant assembly resulting in bundling of nanotubes and loss of emission. Recently, our group[21] and others[22,23] have developed methods to make new SWNT/silica nanocomposites in which some of the intrinsic fluorescence from semiconducting SWNTs is maintained. Our work in particular has focused on making composites under biologically friendly conditions. Specifically, we used a sugar alcohol modified silane molecular precursor, diglyceryl silane (DGS), for the preparation of a silica matrix that can encapsulate fluorescently active SWNTs.[21] DGS-derived silica forms at neutral pH and releases a benign sugar alcohol, glycerol, which makes it effective for encapsulation of SWNTs. However, a significant drop (~66%) in SWNT emission intensity was observed upon encapsulation. Since then, we have been investigating new methods to improve the fluorescence intensity of SWNTs encapsulated in silica so that the PL response observed in the suspension is retained in the composite material. A vapor diffusion sol-gel process,[24] which we refer to here as a Chemical Vapor into Liquids (CViL) sol-gel process, has recently been used to encapsulate fragile assemblies and systems like lipid micelles and bacterial cells without disruption. In addition, the properties (i.e., gelation time and porosity) of the silica gel prepared by the CViL process can be controlled by addition of salt, vapor exposure time, etc.[24] For these reasons, we anticipated that this new approach could be used to prepare SWNT/silica composite materials with minimal disruption to the surfactant micelle adsorbed on the SWNT surface, which should also minimize the changes in emission properties of encapsulated SWNTs.

Here we show that SWNTs can be encapsulated into a silica gel by the CViL process with virtually no loss of SWNT PL intensity. Further, addition of glycerol to SWNT suspensions prior to performing the CViL process allows one to remove water from the nanocomposite gels to prepare a freestanding fluorescent SWNT/silica xerogels. The silica encapsulation process described here has several advantages, including fast preparation and high SWNT loading concentration, over other encapsulation methods used to prepare SWNT/silica nanocomposites that maintain SWNT PL properties. The role of various surfactant systems in determining emission intensity in the silica composites was investigated as a function of environmental conditions, i.e., pH, silica encapsulation parameters, and addition of aromatic molecules known to quench SWNT PL.[5,9,25] We demonstrate that the properties of the CViL-generated silica gels allow the suspension behaviors to be directly translated to the PL response of the gel-encapsulated SWNTs as well. The data presented here shows that SWNT/surfactant systems with higher emission intensities have emission signals that are less susceptible to changes in the surrounding environment. Thus, a balance between the fluorescence intensity of the SWNTs and their sensitivity to changes in the local environment should be considered when utilizing the PL signal from surfactant/SWNT assemblies for new sensing applications.

**RESULTS AND DISCUSSION**

The photophysical properties of SWNTs interacting with either sodium dodecyl sulfate (SDS) or sodium deoxycholate (DOC) were investigated in aqueous solutions and encapsulated in a silica matrix. Although these two surfactants are commonly used to isolate SWNTs, they exhibit distinctly different PL properties under changing conditions. As seen in Figure 1, near-infrared emission from DOC-wrapped SWNTs (Figure 1A, black line) was found to be ~5× brighter than a comparable concentration of SDS-wrapped SWNTs (Figure 1B, black line). The higher PL intensity for the DOC/SWNT suspension may be attributed to the structure of the DOC surrounding assembly, which arises from the amphiphilic nature of the DOC molecules that possess functional groups allowing intermolecular hydrogen bonding. The DOC molecules organize around the SWNTs to maximize the hydrophobic interactions with the SWNT surface, while intermolecular hydrogen bonding results in the formation of a rigid and stable assembly that protects the SWNT surface. This argument is similar to the one used by Ju *et al.*[13] for SWNTs surrounded by flavin-type molecules. In Ju's work, the authors suggest that the flavin molecules pack more tightly around the SWNTs compared to an equivalent concentration of SDS molecules. This packing efficiency leads to reduced interactions between SWNTs and dissolved oxygen that result in increased PL from the SWNTs. In contrast, the structure of SDS is distinctly different with a flexible, linear hydrocarbon tail that interacts with the SWNT surface and a negatively charged sulfate head group that does not strongly interact with other SDS molecules. The resulting SDS assembly around the SWNTs does not appear to pack as efficiently as the DOC surfactant, leading to an assembly that allows a greater interaction between the SWNTs and chemical species (e.g., dissolved $O_2$, $H^+$, etc.) in the surrounding environment that can quench SWNT PL. Similar isolation from the local environment was observed in single molecule studies where individual (6,5) SWNTs suspended in DOC exhibit a brighter and narrower emission peak compared to sodium 4-dodecylbenzenesulfonic acid (SDBS)-SWNTs.[13] Thus, changing the surfactant is one method to improve isolation of SWNTs from the surrounding environment.

To demonstrate that isolation of DOC/SWNT assemblies from the local environment may be similarly translated into functional composite materials, we encapsulated the surfactant/SWNT assemblies in a silica matrix using a chemical vapor into liquids (CViL) sol-gel process.[24] A schematic of the process used to prepare SWNT/silica nanocomposites is shown in Figure 2. It involves placing an open container of tetramethylorthosilicate (TMOS) and a container of an aqueous suspension of surfactant-wrapped SWNTs inside a larger vessel that was sealed in air and kept at room temperature.[24] Due to the volatile nature of TMOS at room temperature, the precursor quickly fills the atmosphere in the sealed vessel. When the vapors reach the aqueous suspension/air interface they hydrolyze to form silicic acid and methanol. Diffusion of silicic acid into the SWNT aqueous suspension followed by condensation leads to formation of Si-O-Si linkages resulting in a homogeneous silica gel. In most cases, the CViL process was performed for 1-2 hrs after which the SWNT suspension was transferred to a capped vessel of choice where the condensation reaction of the silica precursor molecules continues to form a rigid gel (~0.5 day). The slow rate of transfer of TMOS precursor molecules into the SWNT suspension leads to minimal methanol accumulation in the suspension, which is advantageous because methanol can disrupt the SDS/SWNT assembly leading to SWNT aggregation and loss of emission. To demonstrate the lack of methanol in the as-prepared gels, an FT-IR spectrum of an aqueous solution before and after exposure to TMOS vapor was collected (Figure 2B). There are no significant features around 2900 cm$^{-1}$ where one would expect C-H modes to be present due to methanol or incomplete condensation of TMOS. Both spectra also have a broad band between 4000 and 3000 cm$^{-1}$ that corresponds to the fundamental stretching vibrations of different types of hydroxyl groups (i.e., H-OH and Si-OH). The spectrum taken after exposure to TMOS vapors also shows new bands at 1062, 964, 810 and 445 cm$^{-1}$ that may be assigned to characteristic vibrations of Si-O-Si bridges cross-linking the silicate network.[26]

The CViL process to sol-gel encapsulation of SWNTs was found to be a general method that has significant advantages over previously demonstrated routes to preparing SWNT/silica gels that maintain the fluorescence signal from SWNT aqueous suspensions.[21,23] Specifically, the CViL process can be performed at room temperature, does not require addition of an acid or base catalyst and allows one to control the amount of silica precursor added to the solution to be gelled by controlling exposure time and reaction temperature. In addition, the CViL process can be used to induce gelation on highly concentrated SWNT suspensions. Our previous work with diglycerylsilane required that a dilution be made prior to gelation,[21] while no such dilution is necessary using CViL allowing one to make gels with extremely high concentrations of SWNTs. Typical suspensions used here have a 10 mg/L concentration of SWNTs, although there is no limitation imposed by the process which can potentially be important for future optoelectronic and photovoltaic applications. Because the CViL process can be used to gel aqueous solutions one could also include biomolecules or other fragile systems into the SWNT/silica composite materials. Improvements are also found in attainable intensities and stability against SWNT bundling as described below.

The resulting PL properties of the silica encapsulated surfactant/SWNTs assemblies are shown in Figure 1. Upon encapsulation most of the SWNT PL features arising from distinct nanotube chiralities were maintained regardless of surfactant used. In particular, virtually no change was observed in the emission intensities for the DOC/SWNT silica gels. The high quality PL properties observed at the ensemble level in these gels were also found at the single tube level. Figures 1C and 1D show a PL image containing several individual (6,5) nanotubes captured in the silica gel and the PL spectrum from one of those nanotubes, respectively. Uniform PL intensity is found over the length of the tube and the spectral linewidth is 17 nm (fwhm).

For SDS/SWNTs encapsulated in silica there was, however, an overall decrease in intensity of ~50% while virtually no change was observed for DOC/SWNT silica gels. The PL intensity for the larger diameter tubes in the SDS/SWNT/silica composites, which emit at longer wavelengths, also decreased to a much greater extent (see, for example, the emission peak for the (9,7) SWNTs (starred emission feature at ~1322 nm) in Figure 1B. This behavior is typical of SWNT suspensions that undergo a drop in pH as protons are known to interact more strongly with larger diameter SWNTs causing a preferential bleaching of their emission signals.[5,25] In fact, prior to exposure to TMOS vapor the initial pH for the SDS/SWNT suspensions was measured to be 7, while after exposure to TMOS vapor for 1 hr the pH dropped to 5.5, which is consistent with the photoluminescence data shown in Figure 1B. Interestingly, no similar trend was observed from DOC/SWNT suspensions (Figure 1A), although a similar drop in pH was measured. These data indicate that protons, dissolved oxygen, and other quenching species interact with the SWNTs to a greater extent when surrounded by the more loosely-packed SDS coating, while they are effectively excluded by the more tightly packed DOC assemblies.

We attempted to mitigate the effects a drop in pH has on PL from SDS/SWNT assemblies by buffering the initial suspension with a phosphate buffer (pH 7.4, 0.1 M NaCl). In buffered suspensions less than a 0.5 unit drop in pH was measured after exposure to TMOS vapors. In Figure 3, emission spectra for the buffered SDS/SWNT suspension before (green trace) and after (light blue trace) gelation are shown along with emission data for the unbuffered suspension. Addition of phosphate buffer results in an increase in emission intensity of ~30%. Further, only a 10% drop in emission was observed upon formation of a silica gel around the SDS/SWNTs in buffered suspensions. These data are consistent with a model previously proposed by Niyogi *et al.* in which changes in electrolyte concentration can cause an increase in SDS packing density and a reorganization of SDS molecules at the nanotube surface resulting in enhanced PL and volume changes in SDS/SWNT assemblies.[27] This surfactant re-organization may be attributed to interactions between added cations in the buffer and the negatively charged SDS headgroups. The surfactant reorganization model has been corroborated by monitoring the emission properties of SDS suspensions as a function of salt and surfactant concentration.[28] The data shown in this report demonstrate that buffering protects against pH-induced intensity loss in two ways. The first is the typical buffering against pH change. However, the significant PL intensity increase observed on buffering indicates that the SDS assembly surrounding SWNTs in buffered suspensions behaves more like a DOC assembly, i.e., the SDS assembly is more tightly packed and insulates the SWNTs from the surrounding environment. This result suggests that reordering of the micelle structure may be used to limit access to the SWNT surface by other PL quenchers as well.

*Exposure of nanotubes to organic acceptor molecules.* The SDS and DOC wrapped SWNTs were exposed to small organic electronic acceptor molecules such as azobenzene-type (AB) molecules that have previously been shown to bleach the SWNT photoluminescence signal.[9,12,21] Although changes in the photoluminescence signal have been reported as a function of SWNT chirality, we report the overall integrated photoluminescence intensity over the range of 850 – 1650 nm as a function of time (785 nm excitation, Figure 4). In each experiment, several initial points were collected as a baseline prior to adding the AB. As seen in Figure 4, the unbuffered SDS/SWNT suspension (open, blue squares) responds almost instantaneously with virtually complete bleaching upon addition of AB as reported previously.[9] In the case of silica gels prepared from unbuffered SDS/SWNTs (filled, blue squares), diffusion of AB dominates the PL response. The delayed response in PL loss indicates the AB molecules likely reach the region being interrogated by the excitation source after ~5 min. As time progresses nearly the entire photoluminescence signal from the SWNTs in this sample is bleached, but with significantly increased response time (~240 min). These data indicate that, to some degree, AB gains access to the SWNT surface through dynamic fluctuations in the SDS micelle structure. Encapsulation of the SDS/SWNT structures in the silica lattice limits the ability of the SDS micelle to dynamically reorganize, resulting in the slowed quenching kinetics.

Buffered suspensions of SDS/SWNTs (open, red circles) also respond quickly to AB addition, but only lose ~20% of their overall PL intensity in about 30 min. Thus, our expectation is confirmed that SDS reorganization upon addition of buffer into a more tightly packed assembly at the SWNT surface significantly limits the access of AB to the nanotube surface. As per the above discussion, it is likely that even in this case the SDS molecules still have some freedom of movement that allows for AB penetration through the assembly. For silica gels prepared by encapsulating buffered SDS/SWNTs suspensions (filled, red circles) almost no surfactant reorganization is possible as the SDS molecules are hardened into place upon the added interaction with silica. As a result, little to no bleaching was observed over several hours. Similar results were obtained with DOC/SWNT suspensions (open, black triangle) with <5% of the PL intensity quenched over time. Upon silica gel encapsulation of DOC/SWNTs (filled, black triangles) the SWNTs were virtually insensitive to the addition of AB.

*Raman and photoluminescence from SWNT/silica xerogels.* The SWNT/silica gels described above were also dried out to form silica xerogels. Such SWNT/silica xerogel samples may be useful for photonic applications where the presence of water can cause interference like terahertz spectroscopy. Samples were initially dried in air and then placed in an oven at 80°C for 2 days to remove water from the composites. The silica gels shrink significantly upon drying (typically 90% for samples exposed to TMOS vapor for 1 hr at room temperature) although to a lesser extent as TMOS exposure time is increased. Although it was possible to observe PL signals from these samples, the signal was significantly reduced (~80%) for DOC/SWNT/silica xerogels compared to the initial suspension and almost undetectable when prepared from SDS/SWNT suspensions (data not shown).

Fluorescent silica xerogels with encapsulated SWNTs that maintain PL were, however, prepared by adding glycerol to a DOC/SWNT suspension. The suspension with glycerol was exposed to TMOS vapor for >1 hr, then allowed to gel. The resulting gels were left in air for a week to allow for condensation of the silica framework and evaporation of water resulting in a hard, glass-like material. Again the gels shrunk in size depending on the time exposure to TMOS vapor upon drying. However, because glycerol does not evaporate at room temperature and atmospheric pressure, it remains inside the silica framework and less shrinkage was observed compared to samples without glycerol. Further, as more glycerol was added to the initial suspension the resulting xerogel was still rigid but had a more rubbery consistency. Typical fluorescence and Raman spectra for SWNT/silica xerogels are shown in Figure 5. These SWNT/silica xerogel materials were found to have similar overall integrated PL intensity as that observed from the initial DOC/SWNT suspension (Figure 5A). However, interesting diameter-dependent PL trends were observed; a small decrease in intensity was observed for the smaller diameter SWNTs while the larger diameter SWNTs exhibited a slight increase in intensity. In addition, there is a red shift (~ 3-5 nm) observed that is more pronounced for the larger diameter SWNTs. These changes are presumably due to the changing environment around the SWNTs as they move from water into glycerol upon evaporation. The Raman data shown in Figure 5B are also consistent with the PL spectra. Little SWNT bundling was observed in the xerogels as determined by monitoring the radial breathing mode (RBM) peak at 266 cm$^{-1}$, which is known to gain intensity upon increased nanotube bundling (using 785 nm excitation).[29-31] Similar silica xerogels prepared from SDS/SWNT suspensions with glycerol were found to result in significant loss of PL intensity and nanotube bundling (data not shown). These observations corroborate that the DOC assembly surrounding the SWNTs is a rigid and stable structure that allows preparation of highly fluorescent SWNT composite materials.

**CONCLUSIONS**

In summary, tunability and behavioral differences in photoluminescence and environmental sensitivity of various surfactant-wrapped SWNTs dispersed in solution or encapsulated in silica gels/glasses were demonstrated. Addition of glycerol to the initial SWNT suspension was critical in the preparation of fluorescent xerogels. In general, as the photoluminescence intensity was maximized, the sensitivity of the SWNTs to changes in the surrounding environment was minimized. DOC-wrapped SWNTs were found to have the highest photoluminescence signal for the samples studied here, and were virtually insensitive to changes in their surrounding environment. SDS-wrapped SWNTs in unbuffered suspensions, by contrast, had the weakest emission signal and were most sensitive to changes in environmental conditions such as pH or addition of small PL bleaching molecules. However, upon addition of buffer to SDS/SWNT suspensions, an increase in photoluminescence was observed and the magnitude of the response to environmental changes was diminished. These effects may be explained by SDS surfactant reorganization surrounding the SWNTs. In general, these spectroscopic behaviors also translated to the observed PL response in the silica gels, but notably, interaction of the surfactant with the silica lattice is found to further limit the ability of the surfactant assembly to dynamically reorganize. Thus, conditions were determined which demonstrated a balance between fluorescence intensity and environmental sensitivity for surfactant-wrapped SWNT assemblies. The data should help determine optimal conditions for new sensor platforms utilizing the photoluminescent signal from surfactant-wrapped SWNTs.

**EXPERIMENTAL METHODS**

**Preparation of SWNT Suspensions.** Surfactant isolated SWNT suspensions were prepared from SWNTs synthesized by high-pressure decomposition of carbon monoxide (HiPco).[4] The SWNTs (~40 mg SWNTs in 180 ml $D_2O$) were shear mixed with 1 wt% sodium dodecylsulphate (SDS) or sodium deoxycholoate (DOC) in $D_2O$, followed by ultrasonication and centrifugation at 140,000×g for 4 hrs. This procedure yielded an aqueous suspension of individualized SWNTs functionalized by SDS or DOC molecules.

**Preparation of SWNT/silica nanocomposites.** Two vessels, one containing a surfactant-wrapped SWNT aqueous suspension (~1 mL) and one containing TMOS (0.5 mL, Aldrich, 99+%), were placed in a closed container (diameter = 6.5 cm, height = 2.5cm) for 1-2 hours at room temperature. During this time TMOS vapor fills the sealed container and diffuses into the aqueous suspension. The SWNT suspension containing hydrolyzed TMOS (i.e., $Si(OH)_4$) was transferred to a container where condensation of the silica precursor molecules occurred to form a rigid gel in ~12 hrs. SDS/SWNT suspensions were also buffered at pH 7 using a mixture of sodium phosphate and sodium biphosphate. In this case, the suspension was exposed to TMOS for no more than 1 hr (to prevent gelation in the initial container) prior to transfer to another vessel. All pH measurements of the samples were made prior to gelation using a pH meter. SWNT/silica gels prepared in this way were stable for several months when stored in sealed containers to prevent water evaporation. Preparation of xerogels was accomplished by letting water evaporate from the nanocomposite gels in uncapped containers at room temperature for at least a week followed by heating at 80° C for 2 days. The gels shrink significantly during this time to form a solid SWNTs/silica material that was characterized by fluorescence and Raman spectroscopies. Fluorescent DOC/SWNT xerogels were prepared by adding 5-50 wt. % of glycerol to the initial SWNT suspension prior to TMOS vapor exposure.

**Characterization.** Fluorescence spectra were recorded using a near-IR fluorimeter based on a Thermo Nicolet NXR 9600 FTIR paired with a Ge detector. A Xe arc lamp passed through a monochromator was used as the excitation source. Raman measurements were performed using a laser operating at 785 nm with a liquid nitrogen cooled CCD array using a single grating monochromator. The laser power at the sample was 10-15 mW and Raman spectra were integrated over 10 s. Images and spectra of individual nanotubes were obtained as described previously.[13,16] Briefly, (6,5) nanotubes were resonantly excited using a cw dye laser at their second order excitonic resonance ($S_{22}$ at 567nm). Wide-field images of the near-IR PL from the SWNTs (peak emission at 975 nm) were collected by Si-CCD camera (Micromax, Roper Scientific). For spectral identification, the PL originating from individual SWNTs was sent to a 1D cryogenically cooled CCD (OMA V, Roper Scientific) at the output of a spectrometer.

**Redox Doping and Sensing Studies on SWNTs/Silica Materials.** For doping experiments, an aromatic molecule, 4-amino-1,1-azobenzene-3,4-disulphonic acid (AB, Aldrich) was used. ~30 μl of AB (0.1 M (aq)) was added to SWNT suspensions or encapsulated in silica gels formed in a 4 inch NMR tube. Fluorescence spectra were collected as a function of AB diffusion time. For both the SWNT/SDS and SWNT/DOC suspension (aq) studies, 15 μl of 0.1 M AB (aq) was added to ~1 mL of a SWNTs suspension to study redox doping.

**Acknowledgement.** The authors would like to thank the Department of Energy, Office of Science, Basic Energy Sciences for providing funding for this work. JGD would also like to thank the Los Alamos National Laboratory LDRD Director's Postdoc program for funding his work.

**Figure 1. A**) Fluorescence spectra from a DOC/SWNT aqueous suspension (black line) and DOC/SWNT encapsulated in a silica gel (green line). **B**) Fluorescence spectra from a SDS/SWNT aqueous suspension (pH ~7, black line) and SDS/SWNT encapsulated in a silica gel (pH ~5.5, red line). Emission from the (9,7) SWNT is starred in **B**. For these data, all spectra were collected on unbuffered suspensions. **C**) PL image of individual DOC-wrapped SWNTs encapsulated in a silica gel. **D**) PL spectrum of a single DOC/SWNT shown in **C**.

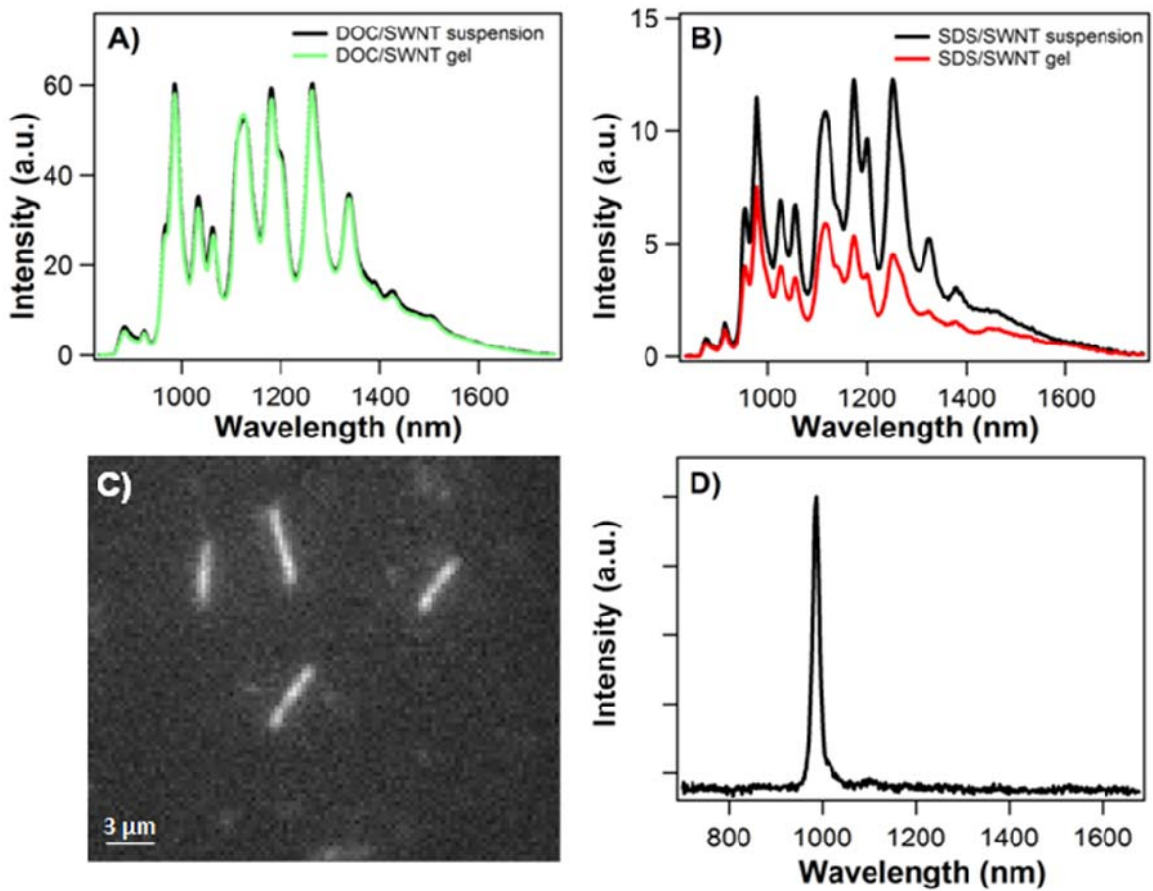

**Figure 2**. **A)** Schematic of the vapor transfer process used to prepare surfactant wrapped SWNT/silica gel nanocomposites. **B)** FT-IR spectra of 10 mM sodium phosphate buffer before exposure (black) and after exposure (red) to TMOS.

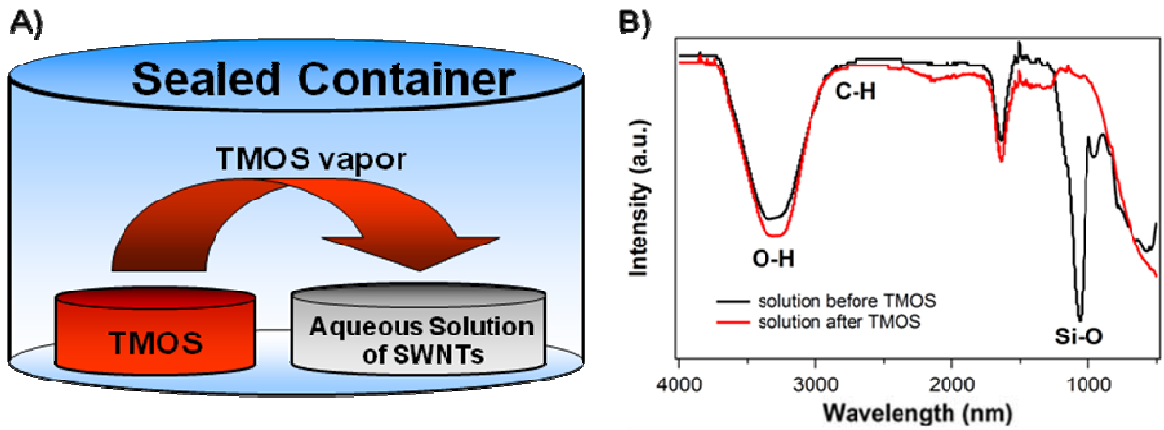

**Figure 3.** Fluorescence spectra of SDS/SWNT suspensions unbuffered (pH ~ 7, black line) and buffered (pH = 7.4, green line), as well as fluorescence spectra of SWNT/silica gels prepared from unbuffered (pH ~ 5.5, red line) and buffered (pH ~ 7, turquoise line) suspensions.

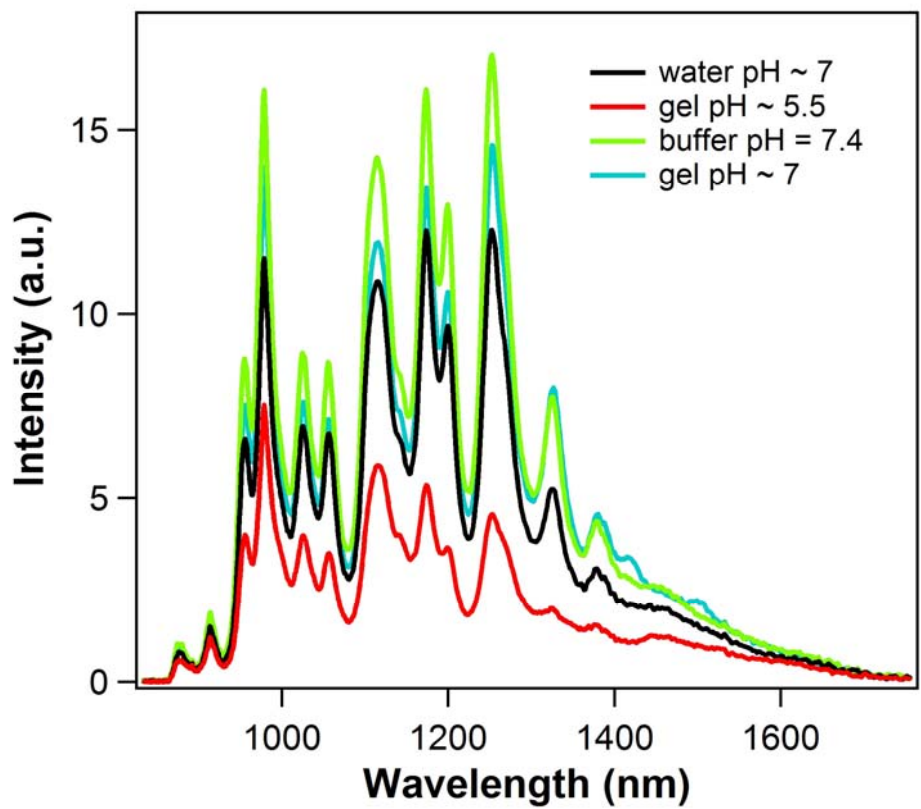

**Figure 4.** Integrated fluorescence intensity (850 – 1650 nm, 785 nm excitation) for surfactant wrapped SWNTs in suspension and silica gels upon exposure to an azobenzene quencher molecule: SDS/SWNTs (unbuffered suspension, □), SDS/SWNTs (unbuffered silica gel, ■), SDS/SWNTs (buffered suspension, ○), SDS/SWNTs (buffered silica gel, ●), DOC/SWNTs (unbuffered suspension, Δ), DOC/SWNTs (unbuffered silica gel, ▲). All buffered suspensions were prepared at pH = 7. The dashed line indicates the time at which the AB complex was added to the suspensions or gels (t = 5min).

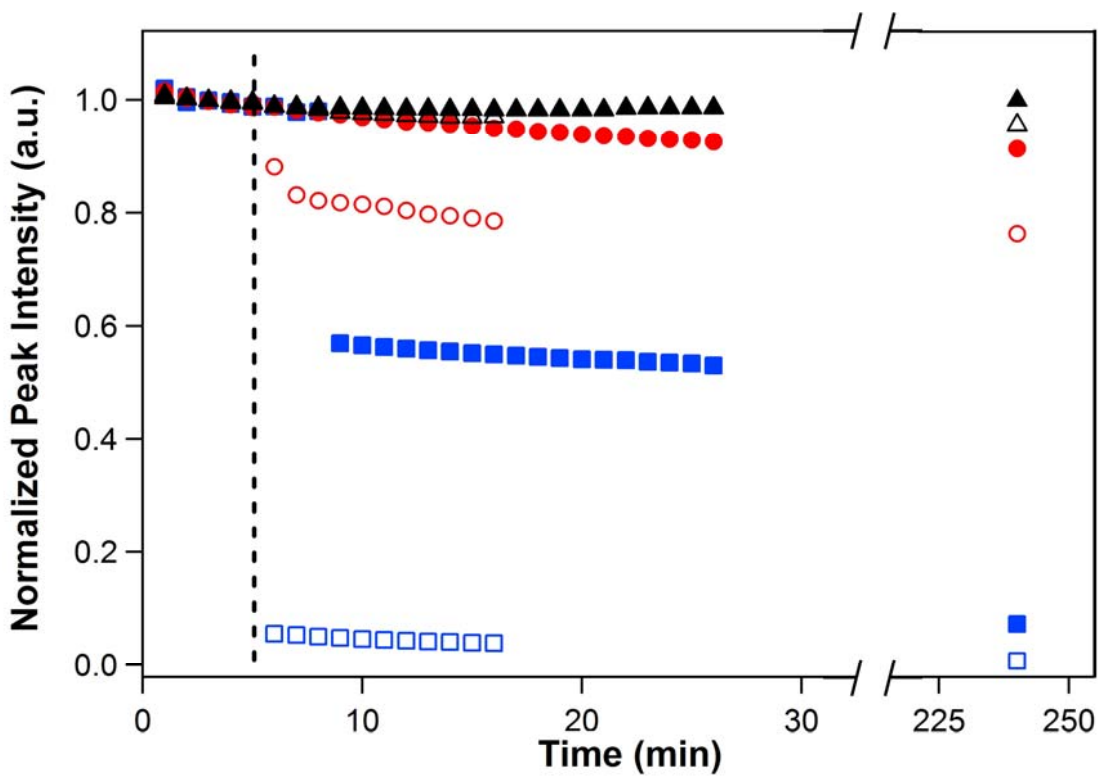

**Figure 5.** **A)** Fluorescence and **B)** Raman spectra collected on an aqueous suspension of DOC/SWNTs with 10% wt. glycerol (black line) and the resulting silica xerogel prepared by vapor diffusion method (green line).

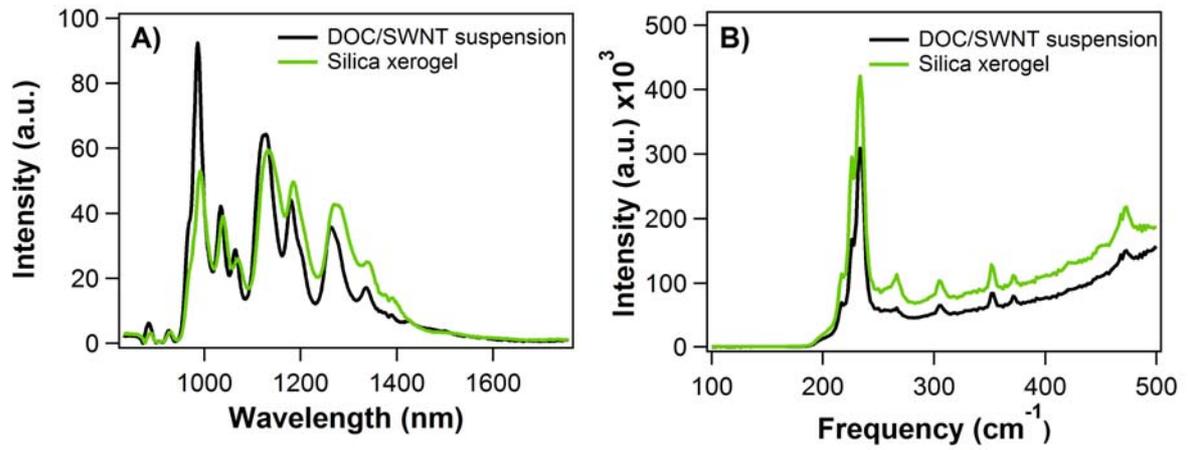